\documentclass{IOS-Book-Article}
\usepackage[T1]{fontenc}
\usepackage[utf8]{inputenc}
\usepackage{graphicx}
\usepackage{mathptmx}
\usepackage{paralist}
\usepackage{epstopdf}
\usepackage{paralist}
\usepackage{tikz}
\usepackage{ctable}
\usepackage{caption,subcaption}
\usepackage{csquotes}
\usepackage{mathtools}
\usepackage{natbib}
\usepackage[textsize=tiny]{todonotes}
\usepackage{url}
\def\hb{\hbox to 10.7 cm{}}

\newcommand{\dara}{\textsf{da\textbar ra}}
\begin{document}

\pagestyle{headings}
\def\thepage{}

\begin{frontmatter}
\title{Identifying and Improving Dataset References in Social Sciences Full Texts}

\author[A,B]{\fnms{Behnam} \snm{Ghavimi}
\thanks{E-mail: Behnam.Ghavimi@gesis.org}},
\author[A]{\fnms{Philipp} \snm{Mayr}
\thanks{E-mail: Philipp.Mayr@gesis.org}},
\author[B]{\fnms{Sahar} \snm{Vahdati}
\thanks{E-mail: s6savahd@uni-bonn.de}}
and
\author[B,C]{\fnms{Christoph} \snm{Lange} 
\thanks{E-mail: langec@cs.uni-bonn.de}}

\address[A]{GESIS – Leibniz Institute for the Social Sciences}
\address[B]{University of Bonn}
\address[C]{Fraunhofer IAIS}
\begin{abstract}

Scientific full text papers are usually stored in separate places than their underlying research datasets.
Authors typically make references to datasets by mentioning them for example by using their titles and the year of publication.	 	
However, in most cases explicit links that would provide readers with direct access to referenced datasets are missing. 
Manually detecting references to datasets in papers is time consuming and requires an expert in the domain of the paper. 
In order to make explicit all links to datasets in papers that have been published already, we suggest and evaluate a semi-automatic approach for finding references to datasets in social sciences papers.
Our approach does not need a corpus of papers (no cold start problem) and it performs well on a small test corpus (gold standard). Our approach achieved an F-measure of 0.84 for identifying references in full texts and an F-measure of 0.83 for finding correct matches of detected references in the {\dara} dataset registry.  
\end{abstract}

\end{frontmatter}
\section{Introduction}
Digital libraries have been growing enormously in recent years.
They provide resources with high metadata quality, easy subject access, and support for retrieving information \citep{Hienert2015}. 
We are specifically interested in scientific full text papers in digital libraries.
Today many papers in the quantitative social sciences make references to datasets. 
However, in most cases the papers do not provide explicit links that would provide readers with direct access to referenced datasets. 

Explicit links from scientific publications to the underlying datasets and vice versa can be useful in multiple use cases.
For example, if a reviewer wants to check the evaluation mentioned in a paper which is performed on a dataset, a link would give him straightforward access to the data, enabling them to check the evaluation.
Or, if other researchers want to perform further analysis on a dataset that was used in a paper, they would be able to do so. 

Today, the majority of papers do not have such direct links to datasets.
While there exist registries that make datasets citable, e.g., by assigning a digital object identifier (DOI) to them, they are usually not integrated with authoring tools.
Therefore, in practice, authors typically cite datasets by \emph{mentioning} them, e.g., using  
combinations of title, abbreviation and year of publication for citing a dataset in the text (see e.g. \citet{Mathiak2015}). 
References to datasets can appear in different places in a paper, as illustrated in figure 1.
It is useful to make all links to datasets explicit in papers that have been published already. 
Manually detecting references to datasets in papers is time consuming and requires an expert in the domain of the paper. 
Detecting dataset references automatically is challenging since in most cases, approaches need a huge corpus of papers as training set. 
We therefore suggest a semi-automatic approach. The system parses full texts very fast and tries to find exact matches without possessing any training set. 
This paper makes the following contributions:
\begin{itemize}
\item a quantitative analysis of typical naming patterns used in the titles of social sciences datasets,
\item a semi-automatic approach for finding references to datasets in social sciences papers with two alternative interactive disambiguation workflows, and
\item an evaluation of the implementation of our approach on a corpus of journal articles.  
\end{itemize}
While a lot of effort has been spent on information extraction in general \citep{Sarawagi2007},
 fewer attempts have focused on the specific task of dataset extraction (see, e.g., \citep{MeiyuLu2012}). 
To refer to the same dataset, different authors often use different names or keywords.
Therefore, simple keyword or name extraction approaches do not solve the problem \citep{Nadeau2007}. 
Each of the references to datasets detected in a paper should be turned into an explicit link, for example by using the DOI of the dataset in a dataset registry. 
In our case, these references should be linked to items in the {\dara} dataset registry. 
\begin{figure}[h]
	\centering
	\includegraphics[width=2.75in]{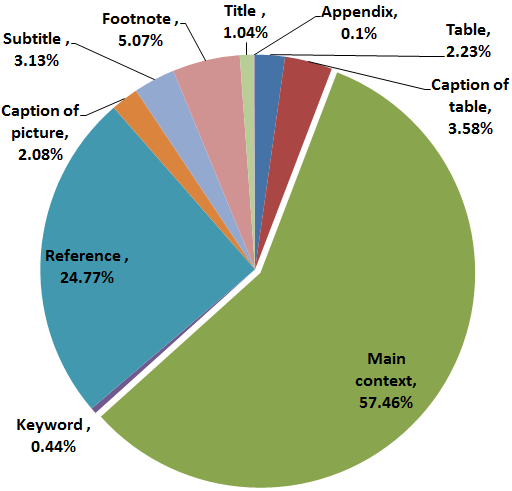}
	\caption{The distribution of dataset references in 15 random mda papers}
	\label{fig:places-example}
\end{figure}
\section{Preliminaries: similarity, ranking and evaluation metrics}
\label{sec:preliminaries}
Our work and other researchers' related work employ certain standard metrics for ranking results of a search query (here: 
a text in a paper that refers to a dataset) over a corpus of documents (here: titles of datasets), and for evaluating the accuracy of information retrieval algorithms.
The following three subsections introduce the definitions of these concepts.
\subsection{Weighting terms in documents using tf-idf}
\label{sec:tfidf}
\emph{Term frequency (tf)} measures the number of occurrences of a given term in a given document or query text \citep{SALTON1988}. 
\emph{Inverse document frequency (idf)} is defined as $\log(N/n)$, where $N$ is the number of all documents in a corpus and $n$ is the number of all documents that contain the given term.
\emph{tf-idf} is defined as the product of \emph{tf} and \emph{idf}. 
When ranking documents that contain a term being searched, tf-idf returns high scores for documents for which the given term is \emph{characteristic}, i.e.\ documents that have many occurrences of the term, while the term has a low occurrence rate in \emph{all} documents of the corpus.
In other words, the tf-idf algorithm assigns a weight to each word in a document, giving high weights to keywords and low weights to frequent words such as stop words. 
\subsection{The cosine similarity metric}
\label{sec:cosine}
A document can be considered as a vector in a vector space, each dimension of which corresponds to one term in the document corpus. 
Such a document vector looks like $d=(t_1w_1,\dots,t_nw_n)$, where $t_i=1$ means that the term $t_i$ exists in the document, and $t_i=0$ means the term is absent in the document. 
tf-idf is one way of computing the weight $w_i$ of terms. 
Search results for a multi-word query in a corpus of documents can be ranked by the \emph{similarity} of each document with the query. Given a query vector $q$ and a document vector $d$, their \emph{cosine similarity} is defined as the cosine of the angle $\theta$ between the two vectors \citep{SALTON1988,ChristopherD1999}, i.e.\ $\operatorname{sim}(q,d)=\cos \theta=\frac{q\cdot d}{\|q\|\,\|d\|}$.
Combining tf-idf and cosine similarity yields a ranked list of documents.
In practice, it may furthermore be necessary to define a cut-off threshold to distinguish documents that are considered to match the query from those that do not \citep{Joachims1997}.
\subsection{Precision and recall of a classifier}
\label{sec:precision-recall}
We aim at implementing a binary classifier that tells us whether or not a certain dataset has been referenced by a paper. 
The algorithm tries to find references of datasets in a paper and then to distinguish a perfect match for each reference in {\dara}. 
The reliability of binary classifiers can be determined by the evaluation metrics of \emph{precision and recall}, and furthermore by F-measure, which combines both.
\section{Related work}
\label{sec:relWork}
While only a few scientific works have been found about the specific task of extracting dataset references from scientific publications, a lot of research has been done on its general foundations including metadata extraction and string similarity algorithms. 
Related work can be divided into three main groups covered by the following subsections.
\subsection{Methods based on the “bag of words” model}
A text can be considered as a set of words and represented as a vector, which indicates absence or presence of a word in the text.
In other words, we can assume a vector space, each of whose dimensions corresponds to one word.
Weights for terms in such vectors need to be adjusted by weighting algorithms such as tf-idf.
\citeauthor{Lee2008} proposed an unsupervised keyword extraction method by using a tf-idf model with some heuristics~\citeyearpar{Lee2008}.
Our approach uses similarity measures for finding a perfect match for each dataset reference in a paper by comparing titles of datasets in a repository to sentences in papers.
Similarity measures such as Matching, Dice 2, Jaccard and Cosine can be applied to a vector representation of a text easily (cf.~\citet{ChristopherD1999}).
The accuracy of algorithms based on such similarity measures can be improved by making them semantics-aware, e.g., representing a set of synonyms as a single vector space dimension.
\subsection{Corpus and Web based methods}
These methods use information about co-occurrence of two texts in documents and are used for measuring semantic similarity of texts.  
\citeauthor{sighal2013} proposed an approach to extract dataset names from articles~\citeyearpar{sighal2013}. 
They employed the NGD algorithm, which estimates the probability of two terms existing separately in a document as well as of their co-occurrence. 
They used two research engines, Google Scholar and Microsoft Academic Search, instead of a local corpus. 
\citeauthor{Schaefer2014} proposed the Normalized Relevance Distance (NRD)~\citeyearpar{Schaefer2014}. 
This metric measures the semantic relatedness of terms. 
NRD extends NGD by using relevance weights of terms. 
The quality of these methods depends on the size of used corpus.
\subsection{Machine learning methods}
Many different machine learning approaches have been employed for extracting metadata, in a few cases also for detecting dataset references. 
For example, \citeauthor{Zhang2006}  
proposed keyword extraction methods based on support vector machines (SVM)~\citeyearpar{Zhang2006}. 
\citeauthor{Kaur2010} conducted a survey on several effective keyword extraction techniques, such as  
conditional random field (CRF) algorithms~\citeyearpar{Kaur2010}. 
\citeauthor{Cui2010} proposed an approach using Hidden Markov Model (HMM) to extract meta data from texts~\citeyearpar{Cui2010}. 
\citeauthor{MeiyuLu2012} used the feature based “Llama” classifier for detecting dataset references in documents~\citeyearpar{MeiyuLu2012}. 
Since there are many different styles for datasets references,
large training sets are necessary 
for these approaches.
\citeauthor{Boland2012} proposed a pattern induction method for extracting dataset references from documents to overcome the necessity of such a large training set~\citeyearpar{Boland2012}.
\section{Data sources}
\label{sec:data}
This section describes the two types of data sources that we use. 
We use full text articles from the journal mda to evaluate the performance of our dataset linking approach, 
and metadata of datasets in the {\dara} dataset registry to identify datasets.
 \subsection{Papers from mda journal}
 
 Methods, data, analyses (mda\footnote{\url{http://www.gesis.org/en/publications/journals/mda/}}) is an open-access journal which publishes research on all questions important to quantitative methods, with a special emphasis on survey methodology. It published research on all aspects of science of surveys, be it on data collection, measurement, or data analysis and statistics. All content of mda is freely available and can be distributed without any restrictions, ensuring the free flow of information that is crucial for scientific progress. We use a random sample of full text articles from mda.
  \subsection{The {\dara} dataset registry}
  
  \subsubsection{{\dara} overview}
  The dataset reference extraction approach presented works for social sciences datasets registered in the {\dara} dataset registry\footnote{\url{http://www.da-ra.de}}.
  {\dara} offers the DOI registration service for social science and economic data. 
  {\dara} makes social science research data referenceable and thus improves its accessibility.
  At the time of this writing, {\dara} holds 428,056 records (datasets, texts, collections, videos or interactive resources); 32,858 of them are datasets. 
  For each dataset, {\dara} provides metadata including title, author, language, and publisher.
  This metadata is exposed to harvesters using a freely accessible API using OAI-PMH (Open Archives Initiative Protocol for Metadata Harvesting).\footnote{\url{http://da-ra.de/oaip/}}
  
  \subsubsection{Analysis of dataset titles in {\dara}}
We analyzed the titles of all datasets in {\dara} and the titles were harvested by using the API of {\dara}.
The analysis shows that about one third of the titles follow a special pattern, which makes them easier to detect in the text of a paper. 
We have identified three such special patterns.
First, there are titles that contain \emph{abbreviations}, by which the dataset is often referred to.
Consider, for example, the full title \enquote{Programme for the International Assessment of Adult Competencies (PIAAC), Cyprus}, which contains the abbreviation \enquote{PIAAC}. 
Secondly, there are \emph{filenames}, as in the example \enquote{Southern Education and Racial Discrimination, 1880-1910: Virginia: VIRGPT2.DAT}, where \enquote{VIRGPT2.DAT} is the name of the dataset file. 
Finally, there are 
\emph{phrases} that explicitly denote the 
existence of datasets in a text, such as \enquote{Exit Poll} or \enquote{Probation Survey}.
\enquote{Czech Exit Poll 1996} is an example of such a dataset title.
Abbreviations and special phrases can be found in about 17 and 19 percentage of the {\dara} dataset titles. The intersection of these two groups is only 1.49 percent.  
Filenames occur in less than one percent of the titles.
\section{A semi-automatic approach for finding dataset references}
\label{sec:approach}
We have realized a semi-automatic approach to find references in a given full text to datasets registered in {\dara}.
The first and last steps of our algorithm require human interaction to improve the accuracy of the result.

\subsection{Step 1: Preparing the dictionary}
\label{sec:preparing-dictionary}
We first prepare a \emph{dictionary} of abbreviations and of special phrases.
\emph{Abbreviations} are initially obtained by applying heuristics to the harvested dataset titles from {\dara}.
The titles are preprocessed automatically before the extraction of abbreviations. 
Titles fully made up of capital letters are removed.
The remaining titles are split at colons (`:'); only the first parts are kept if a colon is found. 
Titles are tokenized (by using nltk, a Python package for natural language processing) and tokens which are not completely in lowercase letters, except for the first letter, not combinations of digits and punctuation marks only, not Roman numerals and not starting with digit, are added to a new list. 
Titles are split based on `-' and `(' symbols.
Afterwards, single tokens before such delimiters will also be added to the list.
Items in the list should only contain the punctuation marks `.', `-', `/', `*' and `\&'. 
Items which contain `/' or `-' and at least one part of which is in lowercase, except for the first letter of each part (e.g. News/ESPN), are removed from the list. 
German and English words and country names are removed from the list. 
Words, fully or partially in capital letters will not be pruned by dictionary (first letter is not included). 
The titles, fully in capital letters are converted into lowercase and tokenized. 
Later, they are pruned by dictionary and then their tokens without definition are added to the list by their original format.
These heuristics detect, for example, \enquote{DAWN} in \enquote{Drug Abuse Warning Network (DAWN), 2008} correctly. 
However, it sometimes detects words that are not references to datasets, such as \enquote{NYPD} in \enquote{New York Police Department (NYPD) Stop, Question, and Frisk Database, 2006}.
As the identification of such false positives is hard to automate, we left this task to a human expert.
They will be added to a list manually and, later, will be removed from the results automatically.

The dictionary of \emph{special phrases} also has to be prepared manually.
A list of terms that refer to datasets such as \enquote{Study} or \enquote{Survey} has been generated manually.
This list contains about 30 items. 
Afterwards, phrases containing these terms were derived by heuristics from titles of actual datasets in {\dara}. 
Three types of phrases are considered here. 
The first type comprises tokens that match items in the dictionary such as \enquote{Singularisierungsstudie}. 
The second category comprises phrases that include \enquote{Survey of} or \enquote{Study of} as a sub phrase, plus one more token that is not a stop word (e.g.\ \enquote{Survey of Hunting}).
The last category is a collection of phrases that contain two tokens, one of which is the dictionary (e.g.\ \enquote{Poll}), and the other one is not a stop word (e.g. \enquote{Freedom Poll}).
The phrase list has finally been verified by a human expert. 
\subsection{Step 2: Detecting dataset references and ranking matching datasets}
\label{sec:detecting-ranking}
Next, we \emph{detect characteristic features} (abbreviations or phrases) of dataset titles in the full text of a given paper.
A paper is split into sentences, and each of these features is searched for in each sentence.
A sentence is split into smaller pieces if one feature repeats inside the sentence more than once, since such a sentence may contain references to different versions of a dataset.
Any phrase identified in this step might correspond to more than one dataset title.
For example, \enquote{ALLBUS}\footnote{Allgemeine Bevölkerungsumfrage der Sozialwissenschaften = German General Social Survey} is an abbreviation for a famous social science dataset, of which more than 150 versions are registered in {\dara}. 
These versions have different titles and for instance, the titles differ by the year of study or the geographic coverage, as in \enquote{EVS - European Values Study 1999 - Italy} or \enquote{European Values Study 2008: Azerbaijan (EVS 2008)}.
 
We solve the problem of identifying the most likely datasets referenced by the text in the paper by \emph{ranking} their titles with a combination of tf-idf and cosine similarity.
In this ranking algorithm, we apply the definitions of section~\ref{sec:preliminaries}, where the query is a candidate dataset reference found in the paper and the documents are the titles of all datasets in {\dara}. 

\subsection{Heuristics to improve ranking in Step 2}
\label{sec:heur-impr-rank}

\begin{figure}[h]
	\centering
	\includegraphics[width=4 in]{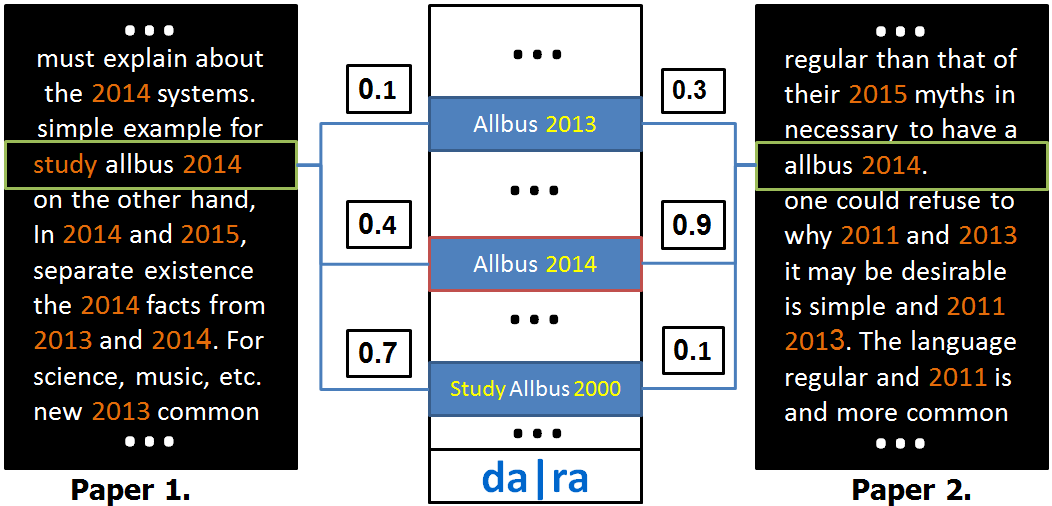}
	\caption{A toy example of cosine similarity, where tf-idf is computed over phrases in two papers and selected titles from {\dara}. The numbers in the figure are not from a real example and just for a demonstration.}
        \label{fig:similarity-example}
\end{figure}

The approach as presented so far computes, for each reference detected in the full text of a paper, tf-idf over the full text of the paper and over the list of the titles of datasets in {\dara} that contain a specific characteristic feature (abbreviation or phrase) detected in the reference. 
While a corpus of papers is typically huge, the size of all {\dara} dataset titles and the size of the full text of an average paper are less than 4~MB each.
Given this limited corpus size, our algorithm may detect some false keywords in a query, thus adversely affecting the result.
For instruction, figure~\ref{fig:similarity-example} illustrates a toy example of this problem.
In paper~1, “2014” repeats many times, whereas “study” occurs once, so tf-idf assigns a high weight to “study” and a low weight to “2014”.
When the query string is “study allbus 2014”, cosine similarity give a higher rank to “Study Allbus 2000” than “Allbus 2014”.  
To address this problem in a better way, our implementation employs some heuristics, including an algorithm that improves the ranking of datasets based on matching years in the candidate strings in the paper and in the dataset titles.
In the example, these heuristics improve the ranking of the “Allbus 2014” dataset when analyzing paper~1.

\subsection{Step 3: Exposing the results to the user, and interactive disambiguation}
\label{sec:expos-results-reus}
The application of our approach supports two workflows by which an expert user can choose the best matches for the datasets cited by a paper from a set of candidates.
The sizes of these sets have been chosen according to the observations we made during the evaluation of the automated step, as explained in section~\ref{sec:eval}.
One workflow works \emph{per reference}: for each reference, five titles of candidate datasets are suggested to the user.
While this workflow supports the user best in getting every reference right, it can be time consuming: each paper in our corpus contains 45 dataset \emph{references} on average, but these only refer to an average number of 3 distinct \emph{datasets}.
The second, alternative, \emph{per-feature workflow} takes advantage of this observation: it works per characteristic feature and suggests, for each feature (which may be common to multiple individual references in the paper) six titles of candidate datasets to the user.

\section{Evaluation}
\label{sec:eval}
 
The calculation of evaluation metrics such as precision, recall and F-measure requires ground truth.
We therefore selected a test corpus of 15 random papers from the 2013 and 2014 issues of the mda journal, 6 in English and 9 in German.
A trained assessor from the InFoLiS II project at GESIS reviewed all papers one by one and identified all references to datasets.
Afterwards, the person attempted to discover a correct match in {\dara} for each detected reference, resulting in a list of datasets per paper.
These lists were used as gold standard, to which we compare the results of our algorithm.  

We decided to divide our evaluation into two steps. 
The first step is about identifying dataset references in papers.
Here, accuracy depends on the quality of the generated dictionaries of abbreviations and special phrases.
These characteristic features (as explained in \ref{sec:detecting-ranking}) are searched by our algorithm in the full texts; detection of any of these features means detection of a reference to a dataset (see row “Detection” in table~\ref{table:eval-results}).
In this phase, if a characteristic feature is identified both in a paper and in the gold standard, it will be labeled as a true positive.
If the feature is in the gold standard but not in our output, it will be labeled as a false negative, or as a false positive in the opposite case. 

The second step of the evaluation matches references detected in papers with items in the {\dara} registry (see row “Matching” in table~\ref{table:eval-results}). 
This evaluation only works on the true positives from the previous step.
The lists of suggested matches for an item, both from the gold standard and from our output, are compared in this step.
An item can have more than one true match since it may occur on its own or in an integrated study (e.g. Allbus 2010 in ALLBUScompact 1980-2012).
In this step, an item will be labeled as a false negative if none of the suggestions for the item in the gold standard appear in our output.
The number of false positives and false negatives are equal in the second step since missing true matches means possessing false positives.
True positives, false positives and false negatives are counted and then used to compute precision and recall. 

The results of our two evaluations are shown in table~\ref{table:eval-results}.
\begin{table}[h!]
	\centering
	\begin{tabular}{c c c c}
		\FL
		Phase of Evaluation & Precision & Recall & F-measure
		\ML
		Detection & 0.91 & 0.77 & 0.84
		\NN
		Matching & 0.83 & 0.83 & 0.83
		\LL
	\end{tabular}
	\caption{Results of the Evaluation }
	\label{table:eval-results}
\end{table}
Our observations in the second evaluation step confirm the choices of set size in the interactive disambiguation workflows. 
In the per-reference matching workflow (as mentioned in \ref{sec:expos-results-reus}), a ranked list of titles of datasets is generated for each of the 45 dataset references (on average in our corpus) in a paper by employing a combination of cosine similarity and tf-idf.
Our observation shows that the correct match among {\dara} dataset titles for each reference detected is in the top 5 items of the ranked list generated by combining cosine similarity and tf-idf for that reference.
Therefore, we adjusted our implementation to only keep the top 5 items of each candidate list for further analysis.
The per-feature matching workflow (as mentioned in \ref{sec:expos-results-reus}) categorizes references by characteristic feature.
For example, in a paper that contains exactly the three detected characteristic features “ALLBUS”, “PIAAC” and “exit poll”, each dataset reference relates to one of these three features.
If we obtain for each such reference the list of top 5 matches as in the per-reference workflow and group these lists per category, we can count the number of occurrences of each dataset title per category.
Now, looking at the dataset titles per category sorted by ascending number of occurrences, 
our results show that the correct matches for the datasets references using a specific characteristic feature were always among the top 6 items.
\section{Conclusion and future work}
\label{sec:future}
We have presented an approach for identifying references to datasets in social sciences papers.
It works in real time and does not require any training dataset.
There are just some manual tasks in the approach such as initially cleaning the dictionary of abbreviations, or making final decisions among multiple candidates suggested for the datasets cited by the given paper.
We have achieved an F-measure of 0.84 for the detection task and an F-measure of 0.83 for finding correct matches for each reference in the gold standard.  
Although the {\dara} registry is large and it is growing fast, there are still many datasets that have not yet been registered there. 
This circumstance will adversely affect the task of detecting references to datasets in papers and matching them to items in {\dara}.
After the evaluation, our observations reveal that {\dara} could cover only 64 percent of datasets in our test corpus. 

Future work will focus on improving the accuracy of detecting references to the datasets supported so far, and on extending the coverage to all datasets.
Accuracy can be improved by better similarity metrics, e.g., taking into account synonyms and further metadata of datasets in addition to the title.
Other algorithms such as identifying the central dataset(s) on which a paper is based can improve the ranked list generated by similarity metrics.
The identification of central dataset(s) is possible after pairing a share of references of datasets in a given paper with titles in {\dara}, and then this identification affects the ranking of rest of the references. 
Coverage can be improved by 
taking into account further datasets, which are not registered in {\dara}.
One promising further source of datasets is OpenAIRE, the Open Access Infrastructure for Research in Europe, which so far covers more than 16,000 datasets from all domains inluding social science but is rapidly growing thanks to the increasing attention paid to open access publishing in the EU.
The OpenAIRE metadata can be consumed via OAI-PMH, or, in an even more straightforward way, as linked data (cf.\ our previous work, \citet{VahdatiEtAl:MappingResearchMetadata15}).
Furthermore, we will enable further reuse scenarios by also exporting RDF from the per-reference matching workflow, using state-of-the-art annotation and provenance ontologies.
For each dataset reference in the paper, we will model the precise position of that reference, and the algorithm's confidence in each possible matching dataset.
In a mid-term perspective, solutions for identifying dataset references in papers that have been published already could be made redundant by a wider adoption of 
standards for properly citing datasets while authoring papers, and corresponding tool support for authors.
\paragraph{Acknowledgements}
This work has been funded by the DFG project “Opening Scholarly Communication in Social Sciences” 
(grant agreements SU 647/19-1 and AU 340/9-1), and by the European Commission under grant agreement 643410
. We thank Katarina Boland from the InFoLiS II project (MA 5334/1-2) for helpful discussions and for generating the gold standard for our evaluation.
\bibliography{biblo} 
\end{document}